\documentclass[prxquantum,twocolumn,notitlepage,nofootinbib,superscriptaddress,longbibliography]{revtex4-2}

\usepackage[dvipsnames]{xcolor}
\usepackage{framed}
\definecolor{shadecolor}{rgb}{0.9,0.9,0.9}

\usepackage{mathtools}
\usepackage{amsmath}
\usepackage[shortlabels]{enumitem}
\usepackage{graphicx,epic,eepic,epsfig,amsmath,latexsym,amssymb,verbatim,color}
 
\usepackage{amsfonts}       % blackboard math symbols
\usepackage{nicefrac}       % compact symbols for 1/2, etc.

\usepackage{amsmath}
\usepackage{bbm}

\usepackage{float}
\usepackage{tikz}
\usetikzlibrary{chains}
\usetikzlibrary{fit}
\usepackage{pgflibraryarrows}		%optional
\usepackage{pgflibrarysnakes}		%optional

\usepackage{epsfig}
\usetikzlibrary{shapes.symbols,patterns} % for source symbols
\usepackage{pgfplots}

\usepackage[strict]{changepage}
\usepackage{hyperref}
\hypersetup{colorlinks=true,citecolor=blue,linkcolor=blue,filecolor=blue,urlcolor=blue,breaklinks=true}

\usepackage[marginal]{footmisc}
\usepackage{url}
\usepackage{theorem}

\newtheorem{proposition}{Proposition}

\newtheorem{theorem}[proposition]{Theorem}

\newtheorem{corollary}[proposition]{Corollary}

\def\squareforqed{\hbox{\rlap{$\sqcap$}$\sqcup$}}
\def\qed{\ifmmode\squareforqed\else{\unskip\nobreak\hfil
\penalty50\hskip1em\null\nobreak\hfil\squareforqed
\parfillskip=0pt\finalhyphendemerits=0\endgraf}\fi}
\def\endenv{\ifmmode\;\else{\unskip\nobreak\hfil
\penalty50\hskip1em\null\nobreak\hfil\;
\parfillskip=0pt\finalhyphendemerits=0\endgraf}\fi}
\newenvironment{proof}{\noindent \textbf{{Proof~} }}{\hfill $\blacksquare$}

\newcounter{remark}

\newcounter{example}

\mathchardef\ordinarycolon\mathcode`\:
\mathcode`\:=\string"8000
\def\vcentcolon{\mathrel{\mathop\ordinarycolon}}
\begingroup \catcode`\:=\active
  \lowercase{\endgroup
  \let :\vcentcolon
  }

\usepackage{cleveref}
\usepackage{graphicx}
\usepackage{xcolor}

\RequirePackage[framemethod=default]{mdframed}
\newmdenv[skipabove=7pt,
skipbelow=7pt,
backgroundcolor=darkblue!15,
innerleftmargin=5pt,
innerrightmargin=5pt,
innertopmargin=5pt,
leftmargin=0cm,
rightmargin=0cm,
innerbottommargin=5pt,
linewidth=1pt]{tBox}

\newmdenv[skipabove=7pt,
skipbelow=7pt,
backgroundcolor=red!15,
innerleftmargin=5pt,
innerrightmargin=5pt,
innertopmargin=5pt,
leftmargin=0cm,
rightmargin=0cm,
innerbottommargin=5pt,
linewidth=1pt]{rBox}

\newmdenv[skipabove=7pt,
skipbelow=7pt,
backgroundcolor=blue2!25,
innerleftmargin=5pt,
innerrightmargin=5pt,
innertopmargin=5pt,
leftmargin=0cm,
rightmargin=0cm,
innerbottommargin=5pt,
linewidth=1pt]{dBox}
\newmdenv[skipabove=7pt,
skipbelow=7pt,
backgroundcolor=darkkblue!15,
innerleftmargin=5pt,
innerrightmargin=5pt,
innertopmargin=5pt,
leftmargin=0cm,
rightmargin=0cm,
innerbottommargin=5pt,
linewidth=1pt]{sBox}
\definecolor{darkblue}{RGB}{0,76,156}
\definecolor{darkkblue}{RGB}{0,0,153}
\definecolor{blue2}{RGB}{102,178,255}
\definecolor{darkred}{RGB}{195,0,0}
\newcommand{\nc}{\newcommand}
\nc{\rnc}{\renewcommand}
\nc{\lbar}[1]{\overline{#1}}
\nc{\bra}[1]{\langle#1|}
\nc{\ket}[1]{|#1\rangle}
\nc{\ketbra}[2]{|#1\rangle\!\langle#2|}
\nc{\braket}[2]{\langle#1|#2\rangle}

\nc{\proj}[1]{| #1\rangle\!\langle #1 |}
\nc{\avg}[1]{\langle#1\rangle}
\nc{\smfrac}[2]{\mbox{$\frac{#1}{#2}$}}
\nc{\tr}{\operatorname{Tr}}
\nc{\sign}{\operatorname{sign}}
\nc{\ox}{\otimes}
\nc{\dg}{\dagger}
\nc{\dn}{\downarrow}
\nc{\cA}{{\cal A}}
\nc{\cB}{{\cal B}}
\nc{\cC}{{\cal C}}
\nc{\cD}{{\cal D}}
\nc{\cE}{{\cal E}}
\nc{\cF}{{\cal F}}
\nc{\cG}{{\cal G}}
\nc{\cH}{{\cal H}}
\nc{\cI}{{\cal I}}
\nc{\cJ}{{\cal J}}
\nc{\cK}{{\cal K}}
\nc{\cL}{{\cal L}}
\nc{\cM}{{\cal M}}
\nc{\cN}{{\cal N}}
\nc{\cO}{{\cal O}}
\nc{\cP}{{\cal P}}
\nc{\cQ}{{\cal Q}}
\nc{\cR}{{\cal R}}
\nc{\cS}{{\cal S}}
\nc{\cT}{{\cal T}}
\nc{\cU}{{\cal U}}
\nc{\cV}{{\cal V}}
\nc{\cX}{{\cal X}}
\nc{\cY}{{\cal Y}}
\nc{\cZ}{{\cal Z}}
\nc{\cW}{{\cal W}}
\nc{\csupp}{{\operatorname{csupp}}}
\nc{\qsupp}{{\operatorname{qsupp}}}
\nc{\var}{{\operatorname{var}}}
\nc{\rar}{\rightarrow}
\nc{\lrar}{\longrightarrow}
\nc{\polylog}{{\operatorname{polylog}}}
\nc{\wt}{{\operatorname{wt}}}
\nc{\av}[1]{{\left\langle {#1} \right\rangle}}
\nc{\supp}{{\operatorname{supp}}}

\nc{\argmin}{{\operatorname{argmin}}}

\def\i{\mathbf{i}}

\def\x{\xi}

\nc{\RR}{{{\mathbb R}}}
\nc{\CC}{{{\mathbb C}}}
\nc{\FF}{{{\mathbb F}}}
\nc{\NN}{{{\mathbb N}}}
\nc{\ZZ}{{{\mathbb Z}}}
\nc{\PP}{{{\mathbb P}}}
\nc{\QQ}{{{\mathbb Q}}}
\nc{\UU}{{{\mathbb U}}}
\nc{\EE}{{{\mathbb E}}}
\nc{\id}{{\operatorname{id}}}

\nc{\CHSH}{{\operatorname{CHSH}}}

\nc{\be}{\begin{equation}}
\nc{\ee}{{\end{equation}}}
\nc{\bea}{\begin{eqnarray}}
\nc{\eea}{\end{eqnarray}}
\nc{\<}{\langle}
\rnc{\>}{\rangle}
\nc{\rU}{\mbox{U}}

\nc{\ob}[1]{#1}

\nc{\SEP}{{\text{\rm SEP}}}
\nc{\NS}{{\text{\rm NS}}}
\nc{\LOCC}{{\text{\rm LOCC}}}
\nc{\PPT}{{\text{\rm PPT}}}
\nc{\EXT}{{\text{\rm EXT}}}
\nc{\Sym}{{\operatorname{Sym}}}

\nc{\ERLO}{{E_{\text{r,LO}}}}
\nc{\ERLOCC}{{E_{\text{r,LOCC}}}}
\nc{\ERPPT}{{E_{\text{r,PPT}}}}
\nc{\ERLOCCinfty}{{E^{\infty}_{\text{r,LOCC}}}}
\nc{\Aram}{{\operatorname{\sf A}}}

\usepackage{tikz}
\usepackage{hyperref}
\hypersetup{colorlinks=true,citecolor=blue,linkcolor=blue,filecolor=blue,urlcolor=blue,breaklinks=true}

\makeatletter
\def\grd@save@target#1{%
  \def\grd@target{#1}}
\def\grd@save@start#1{%
  \def\grd@start{#1}}
\tikzset{
  grid with coordinates/.style={
    to path={%
      \pgfextra{%
        \edef\grd@@target{(\tikztotarget)}%
        \tikz@scan@one@point\grd@save@target\grd@@target\relax
        \edef\grd@@start{(\tikztostart)}%
        \tikz@scan@one@point\grd@save@start\grd@@start\relax
        \draw[minor help lines,magenta] (\tikztostart) grid (\tikztotarget);
        \draw[major help lines] (\tikztostart) grid (\tikztotarget);
        \grd@start
        \pgfmathsetmacro{\grd@xa}{\the\pgf@x/1cm}
        \pgfmathsetmacro{\grd@ya}{\the\pgf@y/1cm}
        \grd@target
        \pgfmathsetmacro{\grd@xb}{\the\pgf@x/1cm}
        \pgfmathsetmacro{\grd@yb}{\the\pgf@y/1cm}
        \pgfmathsetmacro{\grd@xc}{\grd@xa + \pgfkeysvalueof{/tikz/grid with coordinates/major step}}
        \pgfmathsetmacro{\grd@yc}{\grd@ya + \pgfkeysvalueof{/tikz/grid with coordinates/major step}}
        \foreach \x in {\grd@xa,\grd@xc,...,\grd@xb}
        \node[anchor=north] at (\x,\grd@ya) {\pgfmathprintnumber{\x}};
        \foreach \y in {\grd@ya,\grd@yc,...,\grd@yb}
        \node[anchor=east] at (\grd@xa,\y) {\pgfmathprintnumber{\y}};
      }
    }
  },
  minor help lines/.style={
    help lines,
    step=\pgfkeysvalueof{/tikz/grid with coordinates/minor step}
  },
  major help lines/.style={
    help lines,
    line width=\pgfkeysvalueof{/tikz/grid with coordinates/major line width},
    step=\pgfkeysvalueof{/tikz/grid with coordinates/major step}
  },
  grid with coordinates/.cd,
  minor step/.initial=.2,
  major step/.initial=1,
  major line width/.initial=2pt,
}
\makeatother

\usepackage{thmtools}
\usepackage{thm-restate}
\usepackage{etoolbox}
\makeatletter
\def\problem@s{}
\newcounter{problems@cnt}

\newcommand{\allproblems}{\problem@s}
\makeatother

\usepackage{amsmath,amsfonts,amssymb,array,graphicx,mathtools,multirow,bm,times,tcolorbox,relsize,booktabs}
\usepackage[utf8]{inputenc}
\usepackage[T1]{fontenc}
\usepackage{tikz} 
\usetikzlibrary{quantikz2}
\usepackage[ruled, vlined, linesnumbered]{algorithm2e}

\definecolor{colortwo}{rgb}{0.4,0.77,0.17}
\definecolor{colorthree}{rgb}{0.01,0.51,0.93}
\nc{\EPPT}{{E_{\operatorname{PPT}}}}
\nc{\EPPTone}{{E_{\operatorname{PPT}}^{(1)}}}
\nc{\EK}{{E_{\kappa}}}
\allowdisplaybreaks

\begin{document}
\title{Sample-Efficient Estimation of Nonlinear Quantum State Functions}
\author{Hongshun Yao}
\thanks{H. Yao and Y. Liu contributed equally to this work.}
\affiliation{Thrust of Artificial Intelligence, Information Hub,\\
Hong Kong University of Science and Technology (Guangzhou), Guangdong 511453, China}
\author{Yingjian Liu}
\thanks{H. Yao and Y. Liu contributed equally to this work.}
\affiliation{Thrust of Artificial Intelligence, Information Hub,\\
Hong Kong University of Science and Technology (Guangzhou), Guangdong 511453, China}
\affiliation{\small $\langle \textit{aQa}^\textit{L} \rangle$ Applied Quantum Algorithms Leiden, The Netherlands.}
\affiliation{\small Instituut-Lorentz, Universiteit Leiden, P.O. Box 9506, 2300 RA Leiden, The Netherlands.}
\author{Tengxiang Lin}
\affiliation{Thrust of Artificial Intelligence, Information Hub,\\
Hong Kong University of Science and Technology (Guangzhou), Guangdong 511453, China}
\author{Xin Wang}
\email{felixxinwang@hkust-gz.edu.cn}
\affiliation{Thrust of Artificial Intelligence, Information Hub,\\
Hong Kong University of Science and Technology (Guangzhou), Guangdong 511453, China}

\begin{abstract}
Efficient estimation of nonlinear functions of quantum states is crucial for various key tasks in quantum computing, such as entanglement spectroscopy, fidelity estimation, and feature analysis of quantum data. Conventional methods using state tomography and estimating numerous terms of the series expansion are computationally expensive, while alternative approaches based on a purified query oracle impose practical constraints. In this paper, we introduce the quantum state function (QSF) framework by extending the SWAP test via linear combination of unitaries and parameterized quantum circuits. Our framework enables the implementation of arbitrarily normalized degree-$n$ polynomial functions of quantum states with precision $\varepsilon$ using $\mathcal{O}(n/\varepsilon^2)$ copies. We further apply QSF for developing quantum algorithms for fundamental tasks, including entropy, fidelity, and eigenvalue estimations. Specifically, for estimating von Neumann entropy, quantum relative entropy, and quantum state fidelity, where $\kappa$ and $\gamma$ represent the minimal nonzero eigenvalue and normalized factor, respectively, we achieve a sample complexity of $\tilde{\mathcal{O}}(\gamma^2/(\varepsilon^2\kappa))$. Our work establishes a concise and unified paradigm for estimating and realizing nonlinear functions of quantum states, paving the way for the practical processing and analysis of quantum data.
\end{abstract}

\date{\today}
\maketitle
% \tableofcontents

% \newpage
%%%%%%%%%%%%%%%%%%%%%%%%%%%%%%%%%%%%%%%%%%%%%%%%%%%%%%%%%%%%%%%%%%%%%%%%%%%
%%%%%%%%%%%%%%%%%%%%%%%%%%%%%%%%%%%%%%%%%%%%%%%%%%%%%%%%%%%%%%%%%%%%%%%%%%%
% \section{Introduction}\label{sec:intro}
%
\section{Introduction}
Quantum mechanics introduces inherent challenges in extracting complex nonlinear features from quantum states on a quantum computer~\cite{dirac2001lectures}. These nonlinear properties, such as entropy and state fidelity, are crucial in quantum information processing. Such nonlinear characteristics can be formulated as function mappings from quantum states to real numbers, referred to as nonlinear functions of quantum states.

A conventional approach to extracting information from quantum systems is quantum state tomography (QST), which reconstructs the full quantum state via ensemble measurements~\cite{vogel1989determination}. However, QST’s exponential resource scaling renders it impractical for large-scale applications~\cite{cramer2010efficient, christandl2012reliable}. More recently, classical shadow tomography has emerged as an alternative, enabling the estimation of various quantum properties with significantly fewer measurements~\cite{huang2020predicting,huang2022learning}. While classical shadows balance resource efficiency and information retrieval, their predictive power remains limited due to inherent constraints in classical post-processing~\cite{huang2022learning}.

Advanced methods~\cite{zhou2024hybrid} have combined classical shadow tomography with the generalized SWAP test, establishing hybrid quantum-classical frameworks for nonlinear function estimation. Similarly, other approaches~\cite{wang2023quantum} rely on resource-intensive classical processing, such as polynomial coefficient sampling, to estimate nonlinear quantities like $\tr(\rho^k)$ or $\tr(\rho^k\cos(\rho t))$, often assuming a fixed rank for the quantum states~\cite{shin2024rank}. These methods impose substantial sampling complexity and computational overhead, limiting their scalability.

Advancements in quantum computing have also led to quantum signal processing (QSP) and its extensions, such as quantum singular value transformation (QSVT)~\cite{gilyen2019quantum} and quantum phase processing (QPP)~\cite{QPP}, which enable polynomial transformations of quantum states. While QSVT is well-suited for block-encoded matrices, its implementation is constrained by parity restrictions, whereas QPP facilitates trigonometric polynomial transformations of eigenphases in unitary operators~\cite{childs2017quantum}. Despite their versatility, QSP-based methods~\cite{wang2024new,martyn2024parallel,QPP,gilyen2022improved} generally require purified quantum query access, a restrictive assumption that limits their practical applicability.

In this work, we introduce Quantum State Function (QSF), a novel algorithmic framework that circumvents these limitations by leveraging linear combination of unitaries (LCU)~\cite{childs2012hamiltonian,loaiza2023reducing}, the generalized SWAP test~\cite{ekert2002direct}, and parameterized quantum circuits (PQCs)~\cite{benedetti2019parameterized, lubasch2020variational}. Distinct from QSP-based methods, which assume purified quantum query access, our approach operates directly on multiple identical copies of quantum states, thereby significantly broadening its applicability. Specifically, we prove that any degree-$n$ polynomial function of an arbitrary-dimensional quantum state can be computed with precision $\varepsilon$, requiring a sample complexity of $\mathcal{O}(\frac{n}{\varepsilon^2})$. 

Furthermore, we demonstrate the versatility of QSF in estimating quantum entropies and state fidelity, achieving sample complexity of $\tilde{\mathcal{O}}(\frac{\gamma^2}{\varepsilon^2\kappa})$, where $\kappa\in\mathbb{R}_+$ and $\gamma$ denotes the smallest nonzero eigenvalue and a normalization factor, respectively. Numerical experiments validate the accuracy and efficiency of our framework.

\begin{figure}[htbp]
    \centering
    \resizebox{\columnwidth}{!}{
        \begin{quantikz}
            \lstick{$A^\prime\ \ \ \ket{0}$}& & & & \gate{R_y(\theta_1)} & \ctrl{1} & \ \ldots\ & \gate{R_y(\theta_j)}\gategroup[3,steps=2,style={dashed,rounded corners, inner xsep=2pt, inner ysep=2pt},background,label style={label position=north east,anchor=east}]{$C^j-U_j$} & \ctrl{1} & \ \ldots\ & \meter{X}\\
            \lstick{$A\ \ \ \ \ket{0}$} & \qwbundle{\log_2 n}& & \gate{V} & \ctrl{-1} & \ctrl{1} & \ \ldots\  & \ctrl{-1} & \ctrl{1} & \ \ldots\ & \qw\\ 
            \lstick{$B\ \ \rho^{\otimes n}$} & \qwbundle{}& & & & \gate{P_1} & \ \ldots\  & & \gate{P_j} & \ \ldots\ &
        \end{quantikz}
    }
    \caption{Quantum circuit for nonlinear transformation of quantum states. Systems $A^\prime$ and $A$ are initialized into $\ket{0}^{\ox m}$, while the input  $\rho^{\otimes n}$ resides on system $B$. The state preparation circuit $V$ is designed to generate a $\log n$-qubit state $\sum_{j=1}^{n}\sqrt{|\alpha_j|}\ket{j-1}$. $C^j\mbox{-}U_j$ represents a unitary $U_{j}$ controlled by the basis $\ket{j}$ in system $A$. Each $U_j$ comprises a single-qubit rotation $R_y(\theta_j)$ applied to system $A'$, followed by cycle permutations $P_j$ that are jointly controlled by $A'$ and $A$. The rotation angles are set to be $\theta_j:= \text{sign}(\alpha_j)\pi/2$. Then, measuring system $A^\prime$ under Pauli-$X$, one obtains the target function value $f_n(\rho)$. }
    \label{fig:polynomial_transform_general}
\end{figure}
%%%%%%%%%%%%%%%%%%%%%%%%%%%%%%%%%%%%%%%%%%%%%%%%%%%%%%%%%%%%%%%%%%%%%%%%%
%%%%%%%%%%%%%%%%%%%%%%%%%%%%%%%%%%%%%%%%%%%%%%%%%%%%%%%%%%%%%%%%%%%%%%%%%
% \section{Main results}\label{sec:main_results}
%%%%%%%%%%%%%%%%%%%%%%%%%%%%%%%%%%%%%%%%%%%%%%%%%%%%%%%%%%%%%%%%%%%%%%%%%
% \subsection{Quantum state function (QSF)}\label{sec:quantum algorithm} 
\section{The Quantum State Function Framework}

The core of quantum algorithms lies in their ability to manipulate quantum states through unitary transformations and extract meaningful information via measurements. Mathematically, this process can be characterized as a function mapping quantum states to real-valued quantities of interest 
$f: \mathbb{C}^{d \times d} \to \mathbb{R}, \, \rho \mapsto f(\rho)$, where $\mathbb{C}^{d \times d}$ denotes the space of density matrices describing $d$-dimensional quantum states. Such functions encompass fundamental quantities in quantum information theory, including entropies, fidelities, and expectation values of nonlinear observables. A particularly interesting class of polynomial functions for quantum states, defined as $    f_n(\rho) = \sum_{j=1}^{n} \alpha_j \operatorname{Tr}(\rho^j)$, where $\alpha_j \in \mathbb{R}$ are real coefficients. This formulation captures a broad range of physically relevant quantities, such as Rényi entropies and spectral moments. Without loss of generality, we assume the coefficients are normalized so that $\sum_{j=1}^{n} |\alpha_j| = 1$, a convention that simplifies both the implementation and analysis of the associated quantum algorithm. For a sufficiently large polynomial degree $n$, $f_n(\rho)$ provides an arbitrarily accurate approximation of the target nonlinear function $f(\rho)$ up to a real factor $\gamma$, such as Taylor or Chebyshev approximations. Thus, our goal is to develop a quantum framework that efficiently estimates $f_n(\rho)$ while minimizing the required quantum resources and measurement complexity.

% The essence of quantum algorithms lies in their ability to transform quantum states and extract meaningful information through measurements. We formalize this process as a nonlinear function $f: \mathbb{C}^{d\times d} \to \mathbb{R}$, where $\mathbb{C}^{d\times d}$ represents the set of $d$-dimensional quantum states. In this work, we focus on the polynomial function $f_n(\rho)=\sum_{j=1}^n\alpha_j\tr(\rho^j)$ with $\alpha_j \in\mathbb{R}$, and without loss of generality, we assume that it can effectively approximate the nonlinear function $f(\cdot)$ when $n$ is sufficiently large.

% \begin{tcolorbox}
\begin{theorem}\label{thm:polynomial_transform_general}
    For any normalized degree-$n$ polynomial function $f_n(\cdot)$ and quantum state $\rho\in\mathbb{C}^{d\times d}$, there exists a quantum algorithm that estimates $f_n(\rho)$ with precision $\varepsilon$ using $\cO(\frac{n}{\varepsilon^2})$ copies of $\rho$.
\end{theorem}
% \end{tcolorbox}
\begin{proof}
System labels $A', A$ and $B$ are assigned to systems with a single ancilla qubit, $m-1$ ancilla qubits, and $n$ identical copies of $\rho$, as it is shown in Fig~\ref{fig:polynomial_transform_general}. Consider the expectation value of the measurement operator $X$ on system $A'$
\begin{equation}\label{eq:nonlinear_transform_general}
    \langle X \rangle := \tr[U (\ketbra{0}{0}^{\otimes m} \otimes \rho^{\otimes n}) U^\dagger (X_{A'} \ox I_{AB})],
\end{equation}
where $\ketbra{0}{0}^{\otimes m} \otimes \rho^{\otimes n}$ is the initial state with $m = \lceil \log_2 n + 1\rceil$, and $U$ is a quantum circuit designed to extract the polynomial function $f_n(\rho)$, defined as
\begin{equation}\label{eq: U}
    U := \prod_{k=1}^{n} C^k\text{-}U_k \cdot (I_{A^\prime} \otimes V_{A} \otimes I_{B}),
\end{equation}
where $V_A$ prepares the state $\sum_{j=1}^{n} \sqrt{|\alpha_j|} \ket{j-1}$, and each $C^k\text{-}U_k$ consists of a controlled single-qubit rotation $R_y(\theta_k)$ followed by a cycle permutation $P_k$ on $B$. 
Applying $U$ to the initial state yields
\begin{equation}\label{eq:end_V}
    \sum_{j,k=1}^{n} \sqrt{|\alpha_j\alpha_k|} \ketbra{j-1}{k-1}_{A} \otimes U_j (\ketbra{0}{0}_{A'} \otimes \rho^{\otimes n}) U_k^\dagger.
\end{equation}
Taking the expectation of $X_{A^\prime}$, we then obtain
\begin{equation}\label{eq:<x>}
    \langle X \rangle = \sum_{j=1}^{n} |\alpha_j| \sin \theta_j \tr(\rho^j).
\end{equation}
Note that $\theta_j$ is set to be $\text{sign}(\alpha_j)\pi/2$ to recover $ \langle X \rangle = f_n(\rho)$. A random variable $\xi \in \{+, -\}$ is sampled with probabilities $\tr[\sigma_{out} \ketbra{+}{+}]$ and $\tr[\sigma_{out} \ketbra{-}{-}]$, where $\sigma_{out} = \tr_{AB}[U(\ketbra{0}{0}^{\otimes m} \otimes \rho^{\otimes n}) U^\dagger]$. Using Hoeffding’s inequality, the sample mean approximates $\langle X \rangle$ with precision $\varepsilon$ and probability at least $1 - \delta$ with $\cO(\frac{1}{\varepsilon^2} \ln \frac{1}{\delta})$ measurements. Thus, $\cO(\frac{n}{\varepsilon^2} \ln \frac{1}{\delta})$ copies of $\rho$ suffice to estimate $f_n(\rho)$. Omitting constants, this completes the proof.
\end{proof}

The implementation of nonlinear functions necessitates the construction of a state preparation circuit $V$ to generate a $\log n$ qubit state, which is a well-established and crucial task~\cite{bergholm2005quantum,plesch2011quantum,sun2023asymptotically}. This state can be precisely generated using only single-qubit gates and CNOT gates, with depth $\cO(\frac{n}{\log n})$ and width $\cO(n)$~\cite{sun2023asymptotically}. Alternatively, a parameterized quantum circuit (PQC) $V(\bm\beta)$ can be employed to approximate the state $V(\bm\beta)\ket{0}\approx\sum_{j=1}^{n}\sqrt{|\alpha_j|}\ket{j-1}$ with high precision. 
The pre-processing approach demonstrates practical efficacy, as training the PQC with $\log n$ qubits does not suffer Barren Plateau phenomena~\cite{mcclean2018barren,liu2024mitigating} (e.g. for order $n=64$, the width of $V(\bm\beta)$ is only $6$-qubit). Once the trainable parameter $\bm\beta$ is determined, the circuit $V(\bm \beta)$ maintains operational consistency across quantum states. We further propose a variant of the standard QSF framework that circumvents the state preparation circuit $V$ requirement through increased sample complexity, wherein polynomial coefficients are encoded directly into single-qubit rotations $R_y(\theta_j)$. 

Dividing system $B$ into $B_1B_2\cdots B_n$, Theorem~\ref{thm:polynomial_transform_general} implies that the polynomial transformation of state $F_n(\rho):=\sum_{j=1}^n\alpha_j\rho^j$ can also be sketched. To notice it, jointly performing measurements $X \otimes M$ on systems $A'$ and $B_1$ leads to the expectation value $\tr(M F_n(\rho))$. Thus, by introducing a set of randomized measurements $\{M_i\}$ drawn from a tomographically complete ensemble, such as random Pauli or Clifford observables, as in classical shadow tomography~\cite{huang2020predicting, zhou2024hybrid}. This allows us to construct an approximate classical representation on system $B_1$, the classical shadow of $F_n(\rho)$ would be obtained from a limited number of measurements. Thus, the QSF framework not only computes scalar functionals $\tr(F_n(\rho))$, but also permits classical post-processing access to the entire transformed matrix $F_n(\rho)$, circumventing the physical limit of polynomial transformation~\cite{horodecki2003limits}, hence supporting tasks such as spectral feature extraction, quantum kernel construction, and nonlinearly embedded quantum data analysis. 
% \end{remark}
% \end{shaded}

% \YJ{need to rewrite}

The implementation of quantum circuits $U$ allows for recycling of the input state $\rho^{\otimes n}$ during sampling. Following Pauli-$X$ measurement on system $A^\prime$, subsequent Pauli-$Z$ measurements on each qubit of the system $A$ yield outcomes $j\in[0,n-1]$, corresponding to binary eigenvalue strings $\{\pm 1\}^{m-1}$ from the Pauli string $Z^{\otimes (m-1)}$. The probability of obtaining outcome $j$ is $|\alpha_j|$ with the post-measurement state, 
\begin{equation}
   \sigma_{\pm, j}=\dfrac{\rho^{\otimes n}
\pm\frac{1}{2}\sign(\alpha_j)(P_j\rho^{\otimes n}+\rho^{\otimes n}P_j^\dagger)}{1\pm\sign(\alpha_j)\tr\left(\rho^j\right)}, 
\end{equation}
where the cycle permutation operator $P_j$ acts on the first $j$ systems, enabling reuse of the last $n-j$ copies when the outcome on the system $A$ is $j$. Based on Theorem~\ref{thm:polynomial_transform_general}, we introduce the algorithmic framework to realize nonlinear functions of quantum states, as outlined in Algorithm.~\ref{alg:qsf}. 
% Details are provided in Ref~\cite{SM}.
\begin{algorithm}[htbp]
  \SetAlgoLined
  \SetKwData{Left}{left}\SetKwData{This}{this}\SetKwData{Up}{up}
  \SetKwFunction{Union}{Union}\SetKwFunction{FindCompress}{FindCompress}
  \SetKwInOut{Input}{input}\SetKwInOut{Output}{output}
  
  \Input{Normalized coefficients $\{\alpha_j\}^{n}_{j=1}, \, \forall \alpha_j \in \mathbb{R}$, tolerance $\varepsilon$, and  identical copies of $\rho $.}
  
  \Output{Estimation of $f_n(\rho)$}
  % Prepare a $\lceil \log n \rceil$ qubit state via PQCs or exact construction.
  State preparation circuit $V_A$
  
  $\theta_j \leftarrow \text{sign}(\alpha_j)\pi/2, \forall j$
  
  $N \leftarrow 0$, $P_+\leftarrow0$, and $P_{-}\leftarrow 0$
  
  \Repeat{$N\geq \cO\left(\frac{\text{Poly}(n)}{\varepsilon^2}\right)$}
  {1) Initialize ancillary qubits in systems $A'$ and $A$ with $\ket{0}$ and $\ket{0}^{\ox \lceil \log n \rceil}$.
  
  2) Apply quantum circuit $U$ from Eq~\eqref{eq: U} on systems $A', A$, and $B$.
  
  3) Measure the system $A^\prime$ with Pauli-X operator. If the outcome is $+$ we set $P_+\leftarrow P_+ + 1$, otherwise, $P_-\leftarrow P_- + 1$.
  
  4) Measure each qubit in system $A$ with Pauli-$Z$ operators. For outcome $j$, collect in system $B$ from the $n-j+1$-th subsystem to the $n$-th subsystem to recycle $\rho^{\ox n-j}$.
  
  5) $N\leftarrow N+1$
  }
  \KwRet $\hat{f}_n(\rho):=\frac{P_+-P_-}{N}$
  \caption{Quantum State Function (QSF)}
  \label{alg:qsf}
\end{algorithm}
% \textit{Algorithm.---}Based on Theorem~\ref{thm:polynomial_transform_general}, we introduce the algorithmic framework to realize nonlinear functions of quantum states, as outlined in Alg.~\ref{alg:qsfe}.

% The QSF framework enables comprehensive evaluation of both polynomial functions $f_n(\rho)$ and polynomial transformations $F_n(\rho)$ on a quantum computer through single-qubit measurements, eliminating the need for individual term estimation and thereby reducing measurement requirements. Furthermore, QSF operates on multiple identical samples, relaxing the input constraints associated with purified query oracles in QSP-based strategies, thus expanding the applicability of quantum algorithms.

%%%%%%%%%%%%%%%%%%%%%%%%%%%%%%%%%%%%%%%%%%%%%%%%%%%%%%%%
%%%%%%%%%%%%%%%%%%%%%%%%%%%%%%%%%%%%%%%%%%%%%%%%%%%%%%%%
% \subsection{Comparison with existing methods}\label{sec:compare}

% \textit{Comparisons.---}
\section{Comparison with existing methods}
To ensure a fair comparison with existing methods for estimating an arbitrary normalized nonlinear quantum state function, we evaluate the sample complexity — the total number of copies or queries of $\rho$ required to estimate a degree-$n$ polynomial function to precision $\varepsilon$. This metric unifies the query complexity of QSP-based frameworks into a single measure, enabling a direct and meaningful assessment across different protocols.

A widely adopted approach for nonlinear quantum function estimation is the generalized SWAP test~\cite{Bruni2004}, which individually estimates each term $\tr(\rho^j)$ before summing them classically. This independent estimation leads to a sample complexity of at least $\mathcal{O}(\frac{n^2}{\varepsilon^2})$ for $\rho \in \mathbb{C}^{d\times d}$. Alternatively, one can utilize quasi-probability sampling method to implement generalized SWAP tests: define a random variable $\xi := \tr(\rho^j)$ with probabilities $\{|\alpha_j|\}_j$, allowing the polynomial function to be expressed as $f_n(\rho) =  \mathbb{E}[\xi]$. The total copy requirement for quasi-probility sampling escalates to $\mathcal{O}(\frac{n}{\varepsilon^4})$~\cite{wang2023quantum}. In contrast, QSF jointly estimates the entire polynomial function, achieving a reduced complexity of $\mathcal{O}(\frac{n}{\varepsilon^2})$. It is also worth mentioning that if we apply the quantum amplitude estimation technique, the dependence on $\varepsilon$ will be reduced to $1/\varepsilon$ rather than $1/\varepsilon^2$. 

Another category of methods originates from QSP-based frameworks~\cite{low2016methodology, gilyen2022improved, martyn2021grand}, which rely on \textit{purified quantum query access} to apply polynomial transformations to the singular values of block-encoded matrices. QPP framework~\cite{QPP} similarly implements polynomial functions on purified access to $\rho$, requiring $\mathcal{O}(\frac{n}{\varepsilon^2})$ copies under the assumption that each oracle query corresponds to one copy of $\rho$. A recent parallel QSP approach~\cite{martyn2024parallel} extends these techniques to accommodate hardware constraints by combining QSP with parallelizing generalized SWAP tests to reduce circuit depth. The requirement of query depth reduces to $\cO(\frac{n}{k})$ and the width of parallel circuits reduces to $\cO(k)$, yet the integrated number of copies and queries per one-shot experiment remains $\cO(n)$. Hence, this reduction comes at the cost of exponentially ($ k \thicksim \cO(1)$) increasing sample consumption due to the necessity of parallel measurements.

In comparison, QSF eliminates the reliance on purified query access and operates directly on multiple identical copies of $\rho$, aligning with standard quantum state manipulation protocols. This makes QSF a practically viable alternative to QSP-based methods, particularly for applications where purified quantum queries are inaccessible or experimentally costly. QSP-based approaches often require nontrivial polynomial fitting to map a target function to the parameters of a quantum circuit, while our QSF framework admits an explicit and one-to-one embedding from polynomial coefficients to circuit parameters. This direct correspondence eliminates the need for function approximation or angle synthesis and allows for straightforward programmability of arbitrary polynomials. The sample complexity of various methods is summarized in Table~\ref{tab:comprasion}.
\begin{table}[t!]
\centering
\begin{tabular}{l|ccc}
\toprule
    \textbf{Algorithms}  & \textbf{Prerequisites} & \textbf{Sample Complexity} \\
\midrule
    Parallel QSP~\cite{martyn2024parallel}  & Queries \& Copies & $\cO\left(2^{\cO(k)}\text{Poly}(n) /\varepsilon^2\right)$ \\
    QPP~\cite{QPP}  & Purified Queries& $\cO\left(n/\varepsilon^2\right)$  \\
    Generalized SWAP Test~\cite{Bruni2004}  & Identical Copies & $\cO(n^2/\varepsilon^2)$ or $\cO(n/\varepsilon^4)$ \\
    QSF (this work) & Identical Copies& $\cO\left(n/\varepsilon^2\right)$ \\
\bottomrule
\end{tabular}
\caption{Comparisons of integrated sample complexity among different approaches for outputting a $\varepsilon$-estimate to the general polynomial function of quantum states. }
\label{tab:comprasion}
\end{table}

\begin{figure*}[t]
    \centering
    \includegraphics[width=0.8\linewidth]{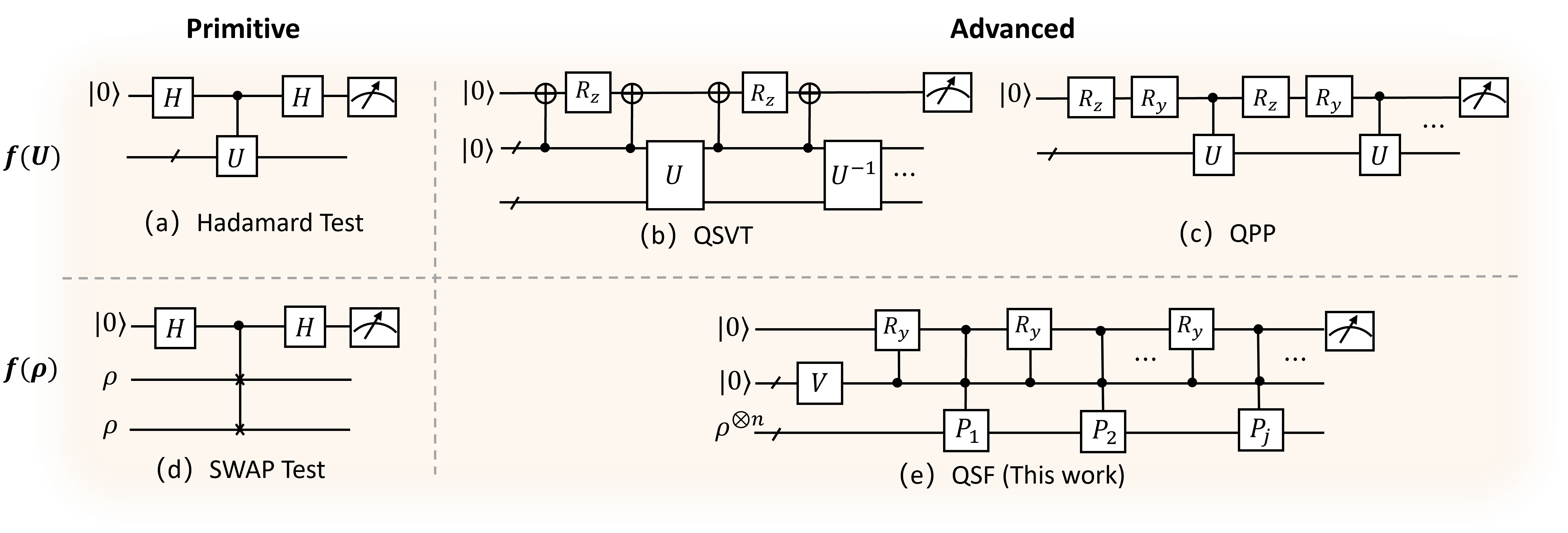}
    \caption{Comparisons of different algorithmic frameworks.  (a), (b), and (c) implement nonlinear functions of unitary operators, $f(U)$, whereas (d) and (e) perform nonlinear functions of quantum states, $f(\rho)$. Technically, (a) and (d) serve as fundamental building blocks of quantum algorithms, while (b), (c), and (e) can be seen as further advancements based on these foundations.}
    \label{fig:quantum_algorithm_comparison}
\end{figure*}

Grounded in quantum algorithmic design, the QSP-based framework can be seen as an extension of the Hadamard test, enabling nonlinear transformations of unitary operators. Consequently, when applying nonlinear transformations to quantum states, these methods rely on encoding the states into unitary operators via block-encoding techniques~\cite{low2019hamiltonian,gilyen2019quantum} or assume direct access to purified query oracles. In contrast, the QSF framework, as a natural extension of the SWAP test, provides a more direct approach to nonlinear quantum state transformations without requiring block encoding or purified quantum queries. This distinction makes QSF particularly well-suited for scenarios where direct manipulation of multiple copies of $\rho$ is feasible, such as distributed quantum computing. A detailed comparison is provided in Fig.~\ref{fig:quantum_algorithm_comparison}.

% \begin{figure}[t]
%     \centering
%     \includegraphics[width=1.0\linewidth]{figs/comparison_pdf.pdf}
%     \caption{Comparisons of different algorithmic frameworks.  (a), (b), and (c) implement nonlinear functions of unitary operators, $f(U)$, whereas (d) and (e) perform nonlinear functions of quantum states, $f(\rho)$. Technically, (a) and (d) serve as fundamental building blocks of quantum algorithms, while (b), (c), and (e) can be seen as further advancements based on these foundations.}
%     \label{fig:quantum_algorithm_comparison}
% \end{figure}

\section{Applications}
The QSF framework provides a unified approach for estimating nonlinear functions of quantum states, with key applications in quantum entropies and distance estimation. In addition, we also propose an effective eigenvalue filter based on QSF framework.

\subsection{Entropy Estimation}

The von Neumann entropy, defined as $\mathcal{S}(\rho) = -\tr(\rho \ln \rho)$, quantifies the information content of a quantum state. Employing polynomial approximations of the logarithm function, QSF framework enables efficient entropy estimation. Specifically, for the polynomial approximation to von Neumann entropy with a normalization factor $\gamma = \sum^n_{j=1}\left|\alpha_j\right|$, our framework promises 
\begin{corollary}\label{co: entropy}
    Given quantum states $\rho\in\mathbb{C}^{d\times d}$ with its minimal nonzero eigenvalue $\kappa \in \mathbb{R}_+$, the QSF framework can approximate $\cS(\rho)$ up to precision $\varepsilon$, with sample complexity $\Tilde{\cO}\left(  \frac{\gamma^2}{\varepsilon^2\kappa}\right)$. 
\end{corollary}

% The proof Corollary~\ref{co: entropy} and detailed circuit implementation of such an estimation can be found in Appendix B. Combining the pre-fixed PQC $V(\bm\beta)$ with the control unitary in the QSF framework, we provide numerical estimation of random qubit state's von Neumann entropy through a degree-$5$ polynomial function $S_5(\rho)$ in Figure.~\ref{fig:entropy_fid}. We simulate the actual single-qubit measurement of system $A'$ for $N_s$ shots to obtain the expectation value $\langle X \rangle$. The shaded bands indicate one standard deviation across 10 independent trials, reflecting the statistical fluctuation. The main plot uses a linear scale to highlight the absolute error and its variance, while the inset log-log view reveals an analogy of power-law convergence behavior with increasing sample size. 

The proof Corollary~\ref{co: entropy} and detailed circuit implementation of such an estimation can be found in Appendix B. It is worth noting that the normalization factor $\gamma$ may grow exponentially with the polynomial degree, as is the case for standard Chebyshev or Taylor expansions of $x \ln x$. To mitigate this potentially prohibitive overhead, one may fix the polynomial degree to a small constant $\cO(1)$ that achieves acceptable precision, and carefully select an appropriate expansion point to reduce the coefficient norm. Alternatively, we provide a variant of QSF framework in Appendix~\ref{apx:quantum_circuits} that relies on the spectral norm instead of 1-norm of the coefficients. This trade-off between approximation accuracy and coefficient norm is crucial to maintaining practical sample complexity in QSF-based estimations.

We present the numerical simulation of von Neumann entropy estimation for a randomly sampled qubit state using a degree-$5$ polynomial approximation $S_5(\rho)$, implemented via a pre-trained two-qubit PQC $V(\bm\beta)$ combined with the controlled unitaries in the QSF circuit. The expectation value $\langle X \rangle$ is obtained from simulated single-qubit measurements on system $A'$ over $N_s$ shots. Figure~\ref{fig:entropy_fid} shows the resulting absolute error across varying measurement shots. The main plot is shown in linear scale to emphasize accuracy, while the inset log-log plot reveals a power-law-like convergence behavior as a function of the number of samples, which aligns with the expected $\cO(\frac{1}{\varepsilon^2})$ measurement cost and converges to the theoretical limit. 

% The shaded region indicates one standard deviation over 10 independent runs, capturing statistical fluctuations. 

Beyond the von Neumann entropy, the QSF framework can also be applied to estimating fractional spectral quantities of the monomial $\tr(\rho^\alpha)$ for $\alpha \in (0,1) $, which are central to many quantum information measures. Notably, the Rényi entropy of order $\alpha > 0$ (and $\alpha \neq 1$) is defined as $\mathcal{S}_\alpha(\rho) := \frac{1}{1-\alpha} \ln \tr(\rho^\alpha),$ serving as a parameterized generalization of the von Neumann entropy. Accurate estimation of $\tr(\rho^\alpha)$ thus enables computation of $\mathcal{S}_\alpha(\rho)$ and plays a crucial role in entanglement detection, quantum thermodynamics, and hypothesis testing. 

To estimate $\tr(\rho^\alpha)$ using QSF, one can construct a polynomial approximation to the function $x^\alpha$ for $\alpha \in (0,1) $ on $x \in [\kappa, 1]$ with degree up to $\cO\left(\frac{1}{\kappa}\log\frac{\log(1/\kappa)}{\varepsilon}\right)$~\cite{gilyen2022improved}, and implement the corresponding estimation via our framework. The following result establishes a sample complexity $\Tilde{\cO}\left(  \frac{\gamma^2}{\varepsilon^2\kappa}\right)$ for estimating $\tr(\rho^\alpha), \alpha \in (0,1) $ in a error tolerance $\varepsilon$. The proof of Corollary~\ref{co: entropy} and detailed description of Rényi entropy, along with the circuit-level implementations, are provided in Appendix~B.

\subsection{Distance Estimation}
Another key application of QSF is the estimation of the distance measures between two quantum states.  Quantum state fidelity $F(\rho, \sigma) = \tr (\sqrt{\rho \sigma}) $ is crucial for quantum verification, benchmarking, and error mitigation. The QSF framework extends naturally to this task by leveraging cyclic permutation techniques to approximate $\tr(\sqrt{\rho \sigma})$ using multiple copies of $\rho$ and $\sigma$. We can further take the fact that $\sqrt{\rho} \sigma \sqrt{\rho}$ shares same eigenvalues with $\rho\sigma$, to simplify it into $\tr ( \sqrt{\rho\sigma})$~\cite{baldwin2023efficiently}. Here, we directly extend the QSF framework to estimate fidelity between two quantum states using a polynomial expansion. To achieve that, we prepare $n$ copies of $\rho$ and $\sigma$, arranged alternately in the sequence $(\rho \ox \sigma )^{\ox n}$. Such a composite operation utilizes the cyclic-permutation-trick, satisfying
\begin{equation}\label{eq: fid}
    \tr[(\rho \ox \sigma )^{\ox n} \cdot P_{2k} ] = \tr[(\rho \sigma)^k], \quad k \leq n, 
\end{equation}
where $P_{2k}$ denotes the cyclic permutation of $2k$ quantum systems. Hence, any normalized polynomial approximation to $f(\rho \sigma)$ up to $n$-th order can be estimated via such an extension to Algorithm~\ref{alg:qsf}. Specifically, we have the following corollary.

\begin{corollary}\label{co: fidelity}
    For quantum states $\rho,\sigma\in\mathbb{C}^{d\times d}$ that the minimal nonzero eigenvalue of $\rho \sigma$ is $\kappa \in \mathbb{R}_+$, the QSF framework can approximate their fidelity $F(\rho, \sigma)$ with precision $\varepsilon$, required sample complexity $\Tilde{\cO}\left(\frac{\gamma^2}{\varepsilon^2\kappa}\right)$. 
\end{corollary}

\begin{figure}[t]
    \centering
    \includegraphics[width=0.9\linewidth]{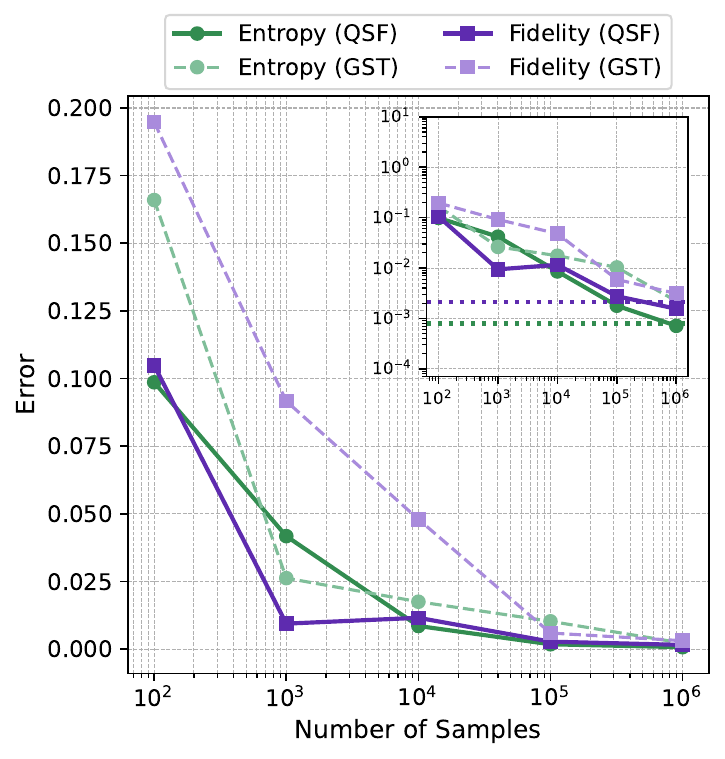}
    \caption{Entropy and fidelity estimation of random states. The solid line, labeled as Entropy (QSF), represents the degree-$5$ averaged estimation obtained using the QSF method, while the purple dashed lines, labeled as Entropy (GST), represent the degree-$4$ estimation from the generalized swap test. For estimating functions of the form $f(\rho) = \sum_j a_j \tr(\rho^j)$ or $\sum_j a_j \tr[(\rho \sigma)^j]$ with generalized swap test circuits, the total measurement cost $N_s$ is distributed across different values of $j$ proportionally to $|a_j| $. The plot presents the resulting absolute error across varying sample sizes and estimation methods. Each estimation is averaged on 10 repeats for trials, and the theoretical errors from polynomial approximation are shown in the log-log subplot with two dotted lines.}
    \label{fig:entropy_fid}
\end{figure}

We show the proof of Corollary~\ref{co: fidelity} in Appendix B. QSF estimates $F(\rho, \sigma)$ with precision $\varepsilon$ using $\Tilde{\mathcal{O}}\left( \frac{\gamma^2}{\varepsilon^2\kappa} \right)$ copies, achieving comparable efficiency to the best-known quantum methods without requiring purified query access. The numerical simulation of quantum state fidelity estimation between two randomly sampled qubit states using a degree-$4$ polynomial approximation $F_5(\rho)$ is presented in Fig.~\ref{fig:entropy_fid}.

Similar to quantum state fidelity, another important distant measure - relative entropy can also be estimated under the QSF framework, also with $\Tilde{\mathcal{O}}\left( \frac{\gamma^2}{\varepsilon^2\kappa} \right)$ sampling complexity.

\subsection{Multivariate function}
The flexibility of the QSF framework covers multivariate functions of quantum states, enabling the estimation of quantities that depend jointly on multiple density matrices. Consider a function of the form \( f(\rho_1, \rho_2, \dots, \rho_m) = \sum_{j} \alpha_j \tr\left( \rho_{i_1}^{k_1} \rho_{i_2}^{k_2} \cdots \rho_{i_r}^{k_r} \right) \), where the indices \( i_1, \dots, i_r \in \{1, \dots, m\} \) and exponents \( k_j \) define the monomials over input states. Such functions arise naturally in the context of quantum channel discrimination~\cite{wang2019resource}, state distinguishability, and multipartite correlation detection~\cite{tura2015nonlocality}. By arranging alternating copies of \( \rho_1, \dots, \rho_m \) and applying generalized cyclic permutation operations, the QSF framework can directly estimate each monomial term $ \tr(\rho_{i_1}^{k_1} \cdots \rho_{i_r}^{k_r}) $, and linearly combine them with coefficients $\alpha_j$ to approximate the full multivariate function.

This multivariate extension inherits the same sampling guarantees as the univariate case. The ability to efficiently estimate such nonlinear correlations underscores the versatility of QSF in broader quantum information processing tasks involving joint statistics of quantum subsystems.

\subsection{Partition function estimation}
Beyond entropy and Rényi-type quantities, the QSF framework naturally extends to estimating thermal partition functions of the form \( \tr(e^{-\beta \rho}) \), which are associated with quantum thermodynamics and quantum generative models. For instance, this trace quantity appears as the normalization factor in Gibbs states and plays a key role in quantum principal component analysis~\cite{lloyd2014quantum} and Hamiltonian simulation~\cite{kimmel2017hamiltonian}.

To estimate \( \tr(e^{-\beta \rho}) \), one can approximate the exponential function via its truncated Taylor series
\begin{equation}
e^{-\beta x} \approx p_N(x) = \sum_{j=0}^N \alpha_j x^j,\quad \alpha_j = \frac{(-\beta)^j}{j!},
\end{equation}
which converges uniformly over \( x \in [\kappa, 1] \) for a fixed inverse temperature \( \beta \). The corresponding 1-norm of coefficients satisfies \( \gamma := \sum_{j=0}^N |\alpha_j| \leq e^\beta \), which provides a tight bound on the normalization cost that $\gamma$ is eligibly regarded as a constant.

\begin{corollary}\label{co: gibbs}
Let \( \rho \in \mathbb{C}^{d \times d} \) be a quantum state with minimal nonzero eigenvalue \( \kappa \in \mathbb{R}_+ \), and fix inverse temperature \( \beta > 0 \). Then the QSF framework can approximate \( \tr(e^{-\beta \rho}) \) up to precision \( \varepsilon \), with sample complexity \( \Tilde{\mathcal{O}}\left( \frac{e^{2\beta}}{\varepsilon^2 \kappa} \right) \).
\end{corollary}

% This result demonstrates the applicability of QSF to thermal expectation values, where QSF avoids the use of purified query oracles typically required in QSP-based approaches. Unlike block-encoded Hamiltonian methods, QSF directly operates on copies of \( \rho \) and admits explicit polynomial embedding, enabling flexible implementation of thermodynamic quantities under mild spectral assumptions.
In the Hamiltonian simulation application, one can exploit the Jacobi-Anger expansion $e^{i\cos{(z)}t}  = \sum^{\infty}_{k = -\infty} i^k J_k(t)e^{ikz}$ to approximate the $e^{-i\rho t}$ with Chebyshev polynomials on spectrum of $\rho$, where $J_k(t)$ denotes the Bessel function of the first kind and the Hamiltonian is encoded on $\rho$.

\subsection{Eigenvalue Filter}\label{max_eigenvalue}

% We developed a hybrid quantum-classical algorithm for estimating the maximum eigenvalue of $d$-dimensional quantum states, illustrating the QSF's versatility and broad applicability. 

Eigenvalues of quantum states provide insights into the statistical properties of quantum states, which are essential for tasks such as principal component analysis (PCA)~\cite{lloyd2014quantum}, and quantum error mitigation~\cite{cai2023quantum}. The variational quantum state eigensolver (VQSE)~\cite{cerezo2022variational} introduced significantly enhances the efficiency of eigenvalue extraction by requiring only a single copy of the density matrix per iteration, making it feasible for noisy intermediate-scale quantum (NISQ) devices. As quantum technologies advance, efficient eigenvalue estimation methods will be pivotal in optimizing quantum algorithms and enhancing the performance of quantum systems.

We propose a hybrid quantum-classical algorithm to estimate the eigenvalue of a $d$-dimensional quantum state $\rho\in\mathbb{C}^{d\times d}$, based on the QSF framework. Without loss of generality, we focus on the maximal eigenvalue of the state. The key ideal is to approximate the step function, $g_{\beta}(x)=0$ for $x\leq\beta$, otherwise, $g(x)=1$, where $\beta\in(0,1)$ is a threshold at which the function $g_{\beta}$ undergoes a sudden jump. 

Suppose the quantum state $\rho\in\mathbb{C}^{d\times d}$ has eigenvalues $\lambda_1\geq\lambda_2\geq\cdots\geq\lambda_d$. Then, one can find that the QSF framework achieves the following value based on a straightforward derivation:
\begin{equation}
    \begin{aligned}
        \sum_{j=1}^n\alpha_j\tr(\rho^j)=\sum_{j=1}^n\alpha_j\sum_{k=1}^d\lambda_k^j=\sum_{k=1}^d\sum_{j=1}^n\alpha_j\lambda_k^j.
    \end{aligned}
\end{equation}
We denote $g_{n,\alpha}(\lambda_k):=\sum_{j=1}^n\alpha_j\lambda_k^j$ as the polynomial approximation of the step function $g_{\beta}$, and notice that the output of our QSF framework can be used as an indication based on the fact that $\sum_{k=1}^dg_{\beta}(\lambda_k)=0$ for $\beta>\lambda_1$, otherwise, it is larger than zero. Specifically, we provide Alg.~\ref{alg:maximal_eigenvalue} for estimating the maximal eigenvalue of the quantum state $\rho$.

According to Theorem~\ref{thm:polynomial_transform_general}, one can find that the step function $g_{\beta}$ with the threshold $\beta\in(0,1)$ can be approximated by our QSF framework with polynomial sample complexity, yielding an approximation error zone with radius $\delta^\prime$ and a tolerance error $\epsilon$. Finally, the threshold $\beta$ will fall within the neighborhood of $\lambda_1$, i.e., $\beta\in(\lambda_1-\delta^\prime,\lambda_1+\delta^\prime)$. 
Notice that the binary search for the maximal eigenvalue $\lambda_1$ requires running the QSF framework logarithmically many times, which implies that the overall sample complexity remains polynomial with respect to the degree of the polynomial function.

\begin{algorithm}[t]
  \SetAlgoLined
  \SetKwData{Left}{left}\SetKwData{Right}{right}\SetKwData{This}{this}\SetKwData{Up}{up}

  \SetKwFunction{Union}{Union}\SetKwFunction{FindCompress}{FindCompress}

  \SetKwInOut{Input}{input}\SetKwInOut{Output}{output}

  \Input{A tolerance $\epsilon$, and $\cO\left(\frac{\text{Poly}(n)}{\varepsilon^2} \right)$ copies of dimensional state $\rho\in\mathbb{C}^{d\times d}$ with certain parameters $\varepsilon$}

  \Output{Estimation of maximal eigenvalue of the state $\rho$}
  
  Initialize the boundaries $\beta_{\text{left}}=0$ and $\beta_{\text{right}}=1$
  
  \Repeat{$\epsilon \leq o_{\beta} \leq 1-\epsilon$}
  {
    Set $\beta = \frac{\beta_{\text{left}} + \beta_{\text{right}}}{2}$
  
    Apply the QSF framework to estimate the step function $g_{\beta}(\rho)$ with the precision $\varepsilon$. Then, denote the output as $o_{\beta}$. 
  
    \If{$o_{\beta} < \epsilon$}{
      $\beta_{\text{right}} \leftarrow \beta$
    }
  
    \If{$o_{\beta} > 1 - \epsilon$}{
      $\beta_{\text{left}} \leftarrow \beta$
    }
  }
  \KwRet$\beta$
  \caption{Maximal Eigenvalue Estimation}
  \label{alg:maximal_eigenvalue}
\end{algorithm}

% While QSVT-based approaches improve upon classical methods, they rely on purified quantum queries, increasing experimental overhead. In contrast, QSF maintains the same asymptotic efficiency while using only standard quantum copies, making it a more practical alternative for real-world implementations. Further details of the comparisons and numerical experiments validating these results are presented in~\cite{SM}, demonstrating the fast convergence of QSF in estimating both entropy and fidelity.

%%%%%%%%%%%%%%%%%%%%%%%%%%%%%%%%%%%%%%%%%%%%%%%%%%%%%%%%%%%%%%%%%%%%%%%%%%%%%%%%%%%%%%%%%%%%%
%%%%%%%%%%%%%%%%%%%%%%%%%%%%%%%%%%%%%%%%%%%%%%%%%%%%%%%%%%%%%%%%%%%%%%%%%%%%%%%%%%%%%%%%%%%%%
% \textit{Conclusions.---}
\section{Conclusions}
We have introduced the Quantum State Function (QSF) framework, a unified approach for implementing nonlinear functions of quantum states. For any degree-$n$ polynomial functions, the QSF framework achieves approximation with precision $\varepsilon$ while requiring only $\mathcal{O}(\frac{n}{\varepsilon^2})$ copies of quantum states. This represents a substantial improvement in sample complexity compared to traditional methods that compute sub-terms separately and rely on classical post-processing. Notably, unlike QSP-based frameworks, QSF offers enhanced applicability by eliminating the requirement for a purified query oracle and angle finding. Few copies of the density matrix can be reused at each iteration of the quantum circuit, making QSF a near-term friendly framework. QSF framework's efficiency in fundamental quantum information tasks has been proven, specifically in estimating quantum entropies and state fidelity. Our work complements existing quantum algorithm toolkits: while QSP-based frameworks have evolved to incorporate the Hadamard test procedure for nonlinear transformations of unitary operators, the QSF framework can be regarded as a natural generalization of the SWAP test, specifically designed for handling nonlinear functions of quantum states.

For future work, exploring the connection between our QSF framework and quantum neural networks (QNNs)~\cite{benedetti2019parameterized,yu2022power,yu2023provable} could lead to a new framework for quantum machine learning~\cite{biamonte2017quantum,cerezo2022challenges}. In the QSF framework, parameters in the rotation gate $R_y$ are fixed, but they could be treated as trainable parameters in QNN models. Building on our theoretical exploration, we aim to develop a quantum machine learning framework with enhanced expressive power and interpretability. Additionally, studying nonlinear functions from the perspectives of quantum information and resource theory could be fruitful, as the nonlinear feature $f_n(\rho)$ might reveal inherent properties of $\rho$, such as entanglement, magic, and coherence resources~\cite{fang2020no,regula2021fundamental,hastings2018distillation,wang2020efficiently,fang2018probabilistic,wang2019resource,chitambar2019quantum}. 
% Focusing on specific nonlinear functions may deepen our understanding of static resources. For instance, obtaining the maximum eigenvalue of a quantum state can be achieved by approximating a step function using the QSF framework.
In summary, further exploration of the QSF framework and its applications in quantum information and computation is worthwhile. The codes and data used in this paper are available at ~\cite{liu2024qsf}.
%%%%%%%%%%%%%%%%%%%%%%%%%%%%%%%%%%%%%%%%%%%%%%%%%%%%%%%%%%%%%%%%%%%%%%%%%
%%%%%%%%%%%%%%%%%%%%%%%%%%%%%%%%%%%%%%%%%%%%%%%%%%%%%%%%%%%%%%%%%%%%%%%%%
\section*{Acknowledgement}
We would like to thank Lei Zhang, Guangxi Li and Zhan Yu, for their helpful discussions and comments. This work was partially supported by the National Key R\&D Program of China (Grant No.~2024YFE0102500), the Guangdong Provincial Quantum Science Strategic Initiative (Grant No.~GDZX2403008, GDZX2403001), the Guangdong Provincial Key Lab of Integrated Communication, Sensing and Computation for Ubiquitous Internet of Things (Grant No.~2023B1212010007), the Quantum Science Center of Guangdong-Hong Kong-Macao Greater Bay Area, and the Education Bureau of Guangzhou Municipality.

%%%%%%%%%%%%%%%%%%%%%%%%%%%%%%%%%%%%%%%%%%%%%%%%%%%%%%%%%%%%%%%%%%%%%%%%%
% Bibliography
% \bibliographystyle{alpha}
\bibliography{ref}

%%%%%%%%% SUPPLEMENTAL MATERIAL %%%%%%%%%

% \newpage

\twocolumngrid
\appendix

\setcounter{subsection}{0}
\setcounter{table}{0}
\setcounter{figure}{0}

\vspace{3cm}
% \onecolumngrid
% \vspace{2cm}
\newpage

% \begin{center}
% \Large{\textbf{Appendix} \\ \textbf{Sample-Efficient Estimation of Nonlinear Quantum State Functions
% }}
% \end{center}

\renewcommand{\theequation}{S\arabic{equation}}
% \numberwithin{equation}{section}
% \renewcommand{\thesubsection}{\normalsize{Supplementary Note \arabic{subsection}}}
\renewcommand{\theproposition}{S\arabic{proposition}}
\renewcommand{\thedefinition}{S\arabic{definition}}
\renewcommand{\thefigure}{S\arabic{figure}}
\setcounter{equation}{0}
\setcounter{table}{0}
\setcounter{section}{0}
\setcounter{proposition}{0}
\setcounter{definition}{0}
\setcounter{figure}{0}

\section{State Preparation}\label{apx:quantum_circuits}
As we introduced in the main text, \textbf{Theorem 1} and Algorithm~\ref{alg:qsf} showed the implementation of the circuit for QSF. In this section, we provide details of state preparation. Additionally, a variant framework is also discussed in the following statement. In this work, we utilize the parameterized quantum circuits (PQCs) shown in Fig.~\ref{fig:parameterized_circuit} to prepare the $\log n$ qubits in system $A$. For an unnormalized polynomial function $f_n(\rho)$ with coefficients $\alpha_j\in\mathbb{R}$. Our goal is to prepare the following state
\begin{equation}
    \ket{\phi} := \sum^n_{j=1}\sqrt{\frac{\left|\alpha_j\right|}{\gamma}}\ket{j-1},
\end{equation}
where $\gamma = \sum^n_{j=1}\left|\alpha_j\right|$.  Denote the output state as $\ket{\psi(\bm\beta)}:=V(\bm\beta)\ket{0}^{\otimes \log n}$. Then, one can achieve the state $\ket{\phi}$ by training the PQC $V(\bm\beta)$ with the following cost function: 
\begin{equation}
    \cL(\bm\beta) := 1 - F(\phi, \psi(\bm\beta)).
\end{equation}

% Combining the pre-fixed PQC $V(\bm\beta)$ with the control unitary in the QSF framework, we provide specific circuits for estimating von Neumann entropy in Figure.~\ref{fig:entropy_circuit} and state fidelity in Figure.~\ref{fig:fidelity_circuit}, respectively.

\begin{figure}[t]
    \centering
    \resizebox{0.4\textwidth}{!}{%
            \begin{quantikz}
                &\qwbundle{\log n}&& \gate{V(\bm\beta)} & \qw
        \end{quantikz}=\begin{quantikz}
            & \gate{R_y(\beta_1)}\gategroup[4,steps=4,style={dashed,rounded corners,fill=blue!20, inner xsep=2pt},background,label style={label position=north east,anchor=south}]{$\times L$}& \ctrl{1}   & \qw & \targ{} & \qw \\
            & \gate{R_y(\beta_2)} & \targ{} & \ctrl{2} & \qw& \qw  \\
            \wave&&&&&\\
            & \gate{R_y(\beta_t)} & \qw & \targ{} & \ctrl{-3} & \qw 
        \end{quantikz}
    }
    \caption{The parameterized quantum circuit for preparing a $\log n$-qubit state. Each layer includes single-qubit $R_y$ rotations and \text{CNOT} gates.}
    \label{fig:parameterized_circuit}
\end{figure}

%%%%%%%%%%%%%%%%%%%%%%%%%%%%%%%%%%%%%%%%%%%%%%%%%%%%%%%%%%%%
% \subsection{Variant framework}\label{apx: variant}
It is worth highlighting that the original QSF framework employs variational quantum algorithms to create superposition states. This reliance on classical operations could potentially limit the versatility of the QSF framework, especially when it is utilized as a subroutine in larger quantum algorithms that do not allow for additional classical operations. To address this potential limitation, we propose a modification to the QSF framework. Our adjustment focuses on replacing the PQC with the Hadamard gate. Specifically, we have the following proposition.

\begin{figure*}[htbp]
    \centering
    \resizebox{0.6\textwidth}{!}{%
        \begin{quantikz}
        \lstick{$A^\prime\ \ \ \ket{0}$} & & & \gate{R_y(\theta_1)} & \ctrl{1} & \gate{R_y(\theta_2)} & \ctrl{1} & \ \ldots\ & \meter{}\\
        \lstick{$A\ \ \ \ket{0}$} & \qwbundle{} & \gate{H} & \ctrl{-1} & \ctrl{1} & \ctrl{-1} & \ctrl{1} & \ \ldots\ & \qw\\ 
        \lstick{$B\ \rho^{\otimes n}$} & \qwbundle{} & & & \gate{P_1} & & \gate{P_2} & \ \ldots\ & \qw\\
        \end{quantikz}
    }
    \caption{Variant circuit. For the target nonlinear function $f_n(\rho):=\sum_{j=1}^n\alpha_j\tr(\rho^j)$ with $\alpha_j\in\mathbb{R}$ and $\gamma:=\max\{|\alpha_j|\}$, we set $\theta_j=\arcsin(\alpha_j/\gamma)$, for all $j$.}
    \label{fig:variant_qsfe}
\end{figure*}

% \begin{shaded}
\begin{proposition}\label{prop:variant_circuit}
     For any degree-$n$ normalized polynomial transform $f_{n}(\rho)=\sum_{j=1}^n\alpha_j\tr(\rho^j)$ with $\alpha_j \in\mathbb{R}$, and quantum state $\rho\in\mathbb{C}^{d\times d}$. The circuit shown in Fig~\ref{fig:variant_qsfe} can estimate $f_n(\rho)$ with precision $\varepsilon$, required sample complexity $\cO(\frac{n^3}{\varepsilon^2})$.
\end{proposition}
% \end{shaded}
\begin{proof}
Similar to the proof in Theorem~\ref{thm:polynomial_transform_general}, one can find that after applying the Hadamard gate $H^{\otimes m-1}$ and controlled unitary $C^k\mbox{-}U_k$ gate, we have the output state
% \begin{aligned}\label{eq: CUj}
%     &\ketbra{+}{+}^{\otimes m-1} \otimes \ketbra{0} {0}\otimes\rho^{\otimes n} \xrightarrow{\oplus_{k=1}^nU_k} \\
%     & \dfrac{1}{n} \sum^{n}_{j,k=1} \ketbra{j-1}{k-1} \otimes U_{j} \left( \ketbra{0}{0}\otimes\rho^{\otimes n} \right) U^{\dagger}_{k}.
% \end{aligned}
\begin{equation}\label{eq: CUj}
    \dfrac{1}{n} \sum^{n}_{j,k=1} \ketbra{j-1}{k-1} \otimes U_{j} \left( \ketbra{0}{0}\otimes\rho^{\otimes n} \right) U^{\dagger}_{k}.
\end{equation}
Then, circuit $U$ ends with a local measurement by the Pauli-$X$ operator, returning the expectation value

\begin{align}
   \langle X \rangle  &= \sum^{n}_{j=1} \frac{\sin{\theta_j} \cdot \tr(I_A)}{4n} \tr(\rho^{\otimes n} P^{\dagger}_j + P_j \rho^{\otimes n}) \\
   &= \sum^{n}_{j=1} \frac{\sin{\theta_j}}{n} \tr(\rho^j).
\end{align}

By setting parameters $\theta_j = \arcsin(\alpha_j/\gamma)$, where $\gamma:=\max\{|\alpha_j|\}$, one could obtain the nonlinear transform $f_n(\rho)$ by the expectation value $\langle X \rangle$ and the amplification with $n\gamma$. Furthermore, we aim to estimate the expectation value $\langle X \rangle$. Similar to the proof in Theorem~\ref{thm:polynomial_transform_general}, one can see that $\cO(\frac{n^3}{\varepsilon^2})$ copies of the quantum state $\rho$ are required for estimating the nonlinear function $f_n(\rho)$ with the precision $\varepsilon$. 
\end{proof}

Although the modified QSF demonstrates an increase in sampling complexity, it extends the algorithm's versatility. This enhancement is particularly notable in situations where constraints prohibit the preparation of assumption states or the introduction of excessive classical computational processes.

%%%%%%%%%%%%%%%%%%%%%%%%%%%%%%%%%%%%%%%%%%%%%%%%%%%%%%%%%%%%%%%%%%%%%%%%%%%
%%%%%%%%%%%%%%%%%%%%%%%%%%%%%%%%%%%%%%%%%%%%%%%%%%%%%%%%%%%%%%%%%%%%%%%%%%%
% \section{Applications}
% \begin{center}
% \large{\textbf{S2. Applications} }
% \end{center}
\section{Estimating Entropies}\label{ap: entropy}
The von Neumann entropy extends the concept of Shannon entropy to quantum systems and serves as a fundamental measure of quantum information, quantifying the uncertainty of a quantum state $\rho$ through $\cS(\rho):= -\tr(\rho \ln \rho)$. In this section, we are going to estimate it using our QSF framework. Specifically, we have the following theoretical guarantees that has been shown in the main text

% \begin{tcolorbox}
    
% \end{tcolorbox}

\begin{corollary}\label{co:entropy}
    Given quantum states $\rho\in\mathbb{C}^{d\times d}$ with its minimal nonzero eigenvalue $\kappa \in \mathbb{R}_+$, the QSF framework can approximate $\cS(\rho)$ up to precision $\varepsilon$, with sample complexity $\Tilde{\cO}\left(  \frac{\gamma^2}{\varepsilon^2\kappa}\right)$. 
\end{corollary}
\begin{proof}
Notice that the natural logarithm $\ln(\cdot)$ is applied as a matrix function to the quantum state $\rho$, operating on the eigenvalues $\{\lambda_i\}$ as $\cS(\rho) = -\sum^{r}_{i=1} \lambda_{i} \ln  \lambda_{i}$, with $r$ denoting the rank of $\rho$. According to Corollary 66 in Ref~\cite{gilyen2019quantum} and Theorem 10 in Ref~\cite{QPP}, there is a polynomial $f_{N}(x) \in \mathbb{R}[x]$ such that for all $x\in[\kappa, 1]$
\begin{equation}
    \left|f_{N}(x) - \dfrac{\ln x}{2\ln\kappa}\right| \leq \dfrac{\varepsilon}{4 \ln(1/\kappa)},
\end{equation}
with degree $N =\cO\left(\frac{1}{\kappa}\log\frac{\log(1/\kappa)}{\varepsilon}\right)$. By denoting the normalized polynomial as $\tilde{f}_{N}:=f_N/\gamma$, one can leverage Algorithm~\ref{alg:qsf} to estimate the $\tr(\rho \tilde{f}_{N}(\rho))$ such that 
\begin{equation}
    \left|\hat{f} - \tr(\rho \tilde{f}_{N}(\rho))\right| \leq \frac{\varepsilon}{4\gamma\ln(1/\kappa)},
\end{equation}
in a cost of $\cO(\frac{N \gamma^2}{\varepsilon^2} \log^2(\frac{1}{\kappa}))$ copies of $\rho$. Hence,  $\hat{f}$ approximates the von Neumann entropy $S(\rho) = \tr(\rho \ln \rho) $ up to the multiplication factor $2\gamma\ln\kappa$ as
\begin{align}
    \left|\hat{f} - \frac{S(\rho)}{ 2\gamma\ln\kappa} \right| &\leq \left|\hat{f} -\tr(\rho \tilde{f}_{N}(\rho)) \right| + \left|\tr(\rho \tilde{f}_{N}(\rho)) - \frac{S(\rho)}{ 2\gamma\ln\kappa} \right| \\
    &\leq \frac{\varepsilon}{4\gamma\ln(1/\kappa)} + \frac{1}{\gamma}\left|
    \tr[\rho ( \tilde{f}_{N}(\rho) -  \frac{\ln\rho}{ 2\ln\kappa} )]\right| \\ \label{eq: holder}
    &\leq \frac{\varepsilon}{4\gamma\ln(1/\kappa)} + \frac{1}{\gamma} \| \tilde{f}_{N}(\rho) -  \frac{\ln\rho}{ 2\ln\kappa} \|_{\infty} \\
    &\leq \frac{\varepsilon}{2\gamma\ln(1/\kappa)},
\end{align}
where we utilize Hölder's inequality to obtain the operator norm in Eqn~\eqref{eq: holder}. Equivalently, it is showed that $ |(2\gamma\ln\kappa) \hat{f} - S(\rho)  | \leq \varepsilon$. That it, to estimate von Neumann entropyni in $\varepsilon-$close precision, $\cO(\frac{\gamma^2}{\varepsilon^2 \kappa} \log\frac{\log(1/\kappa)}{\varepsilon} \log^2(\frac{1}{\kappa}))$ copies of $\rho$ are required. By omitting the logarithms, we obtain the sample complexity $\cO(\frac{\gamma^2}{\varepsilon^2 \kappa})$.
\end{proof}

\begin{figure*}[htbp]
    \centering
    \includegraphics[width=0.8\textwidth]{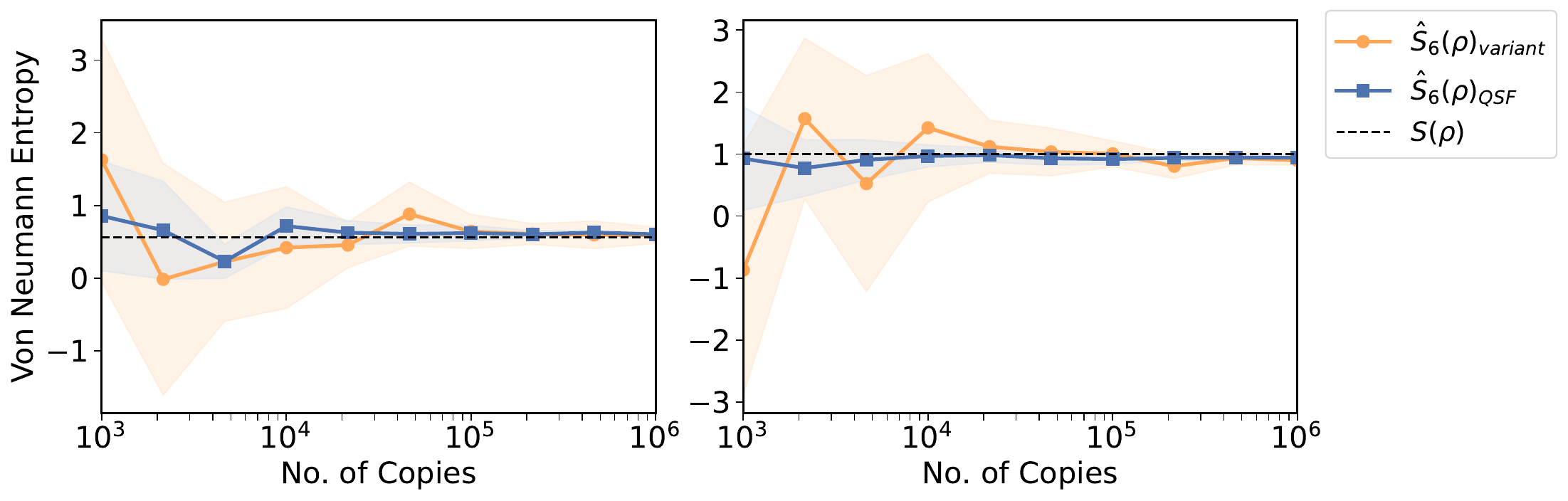}
    \caption{Numerical analysis of von Neumann entropy estimation. (a) A random qubit state. (b) The completely mixed state $I/2$. The dashed black line represents the actual entropy value, denoted as $S(\rho)$. The orange curves, labeled as $\hat{S}_6(\rho)_{\text{variant}}$, represent the degree-6 average estimation obtained using the variant QSF method, while the blue curves, labeled as $\hat{S}_6(\rho)_{\text{QSF}}$, represent the estimation from the standard QSF method. Each estimation is based on 10 repeats for trials, with the shadowed area representing the standard deviation. The specific quantum circuit can be found in Ref.~\cite{SM}.}
    \label{fig:entropy_numerical}
\end{figure*}

The above corollary guarantees the convergence of estimating von Neumann entropy via the QSF framework. Notably, efficient estimation of quantum relative entropy can also be guaranteed under the QSF framework by modifying the input state. We provide the numerical experiments for Algorithm~\ref{alg:qsf} and its variant proposed above. Combining the pre-fixed PQC $V(\bm\beta)$ with the control unitary in the QSF framework, we provide specific circuits for estimating von Neumann entropy through a degree-$6$ polynomial function $S_6(\rho)$ in Fig.~\ref{fig:entropy_circuit}. Numerical simulations are presented in Fig~\ref{fig:entropy_numerical} to estimate the von Neumann entropy of the maximally mixed state and a random quantum state, the simulation is based on Taylor expansions of $S(x)$ on $x_0 = 1$. As shown in Fig.~\ref{fig:entropy_numerical}, the standard QSF framework demonstrates better convergence compared to the variant. The gap arises from an additional factor $\frac{1}{n}$ in the expectation value of the variant. It is seen that the estimation converges to von Neumann entropy at $10^5$ and $10^6$ copies for the standard and variant QSF, respectively. 

\begin{figure*}[htbp]
    \centering
    \resizebox{\textwidth}{!}{%
        \begin{quantikz}
            \lstick{$A^\prime\ \ \ \ \ \ \ket{0}$} & &  & & & & 
            & \gate{R_y(\frac{\pi}{2})} & \gate{R_y(-\frac{\pi}{2})} & \ctrl{1} & \gate{R_y(\frac{\pi}{2})} & \ctrl{1} & \gate{R_y(-\frac{\pi}{2})} & \ctrl{1} & \gate{R_y(\frac{\pi}{2})} & \ctrl{1} & \gate{R_y(-\frac{\pi}{2})} & \ctrl{1} & \meter{X}\\
            \lstick[3]{$A\ \ket{0}^{\otimes 3}$} & & \gate{R_y(\beta_1)} \gategroup[3,steps=4,style={dashed,rounded corners,fill=blue!20, inner xsep=0pt},background,label style={label position=north east,anchor=north west}]{$\times 2$} & \ctrl{1} & & \targ{} &
            & \octrl{-1}    & \octrl{-1}    & \octrl{1} & \octrl{-1}    & \octrl{1} & \octrl{-1}    & \octrl{1} & \ctrl{-1} & \ctrl{1}  & \ctrl{-1}    & \ctrl{1}  & \qw\\
            & & \gate{R_y(\beta_2)} & \targ{} & \ctrl{1} & &
            & \octrl{-1}    & \octrl{-1}    & \octrl{1} & \ctrl{-1}     & \ctrl{1}  & \ctrl{-1} & \ctrl{1}  & \octrl{-1} & \octrl{1} & \octrl{-1}    & \octrl{1} & \qw\\
            & & \gate{R_y(\beta_3)} & & \targ{} & \ctrl{-2} &
            & \octrl{-1}    & \ctrl{-1}     & \ctrl{1}  & \octrl{-1}    & \octrl{1} & \ctrl{-1} & \ctrl{1}    & \octrl{-1} & \octrl{1}    & \ctrl{-1} & \ctrl{1} & \qw\\
            \lstick{$B\ \ \ \ \ \ \rho^{\otimes 6}$} & \qwbundle{} & & & & & & & & \gate{P_2} & & \gate{P_3} & & \gate{P_4} & & \gate{P_5} & & \gate{P_6} & \qw
        \end{quantikz}
    }
    \caption{Quantum circuit for calculating the von Neumann entropy of a quantum state $\rho \in \mathbb{C}^{d \times d}$, using the polynomial function $S_6(\rho) = \sum_{j=1}^6 \alpha_j \tr(\rho^j)$, where $\alpha_j \in \mathbb{R}$. The purple section represents a predetermined PQC $V(\bm\beta)$ designed to generate a 3 qubit state and $P_j$, $j=2,\cdots,6$, denotes the permutation operators.}
    \label{fig:entropy_circuit}
\end{figure*}

A comparison to previous quantum algorithms designed for estimating von Neumann entropy is provided in Table~\ref{tab: comparison entropy}.  Compared to other methods, such as Ref~\cite{wang2023quantum} and Ref~\cite{wang2024new}, QSF maintains a balance between practical applicability and theoretical efficiency, positioning it as a promising approach for entropy estimation.

\begin{table}[H]
\centering
\begin{tabular}{l|ccc}
\toprule
    \textbf{Algorithms}  & \textbf{Prerequisites} & \textbf{Sample Complexity} \\
\midrule
    Ref~\cite{wang2023quantum}  & Identical Copies & $\Tilde{\cO}\left(  1/(\varepsilon^5\kappa^2)\right)$ \\
     QSVT~\cite{wang2024new}  & Purified Query& $\Tilde{\cO}\left( 1/(\varepsilon^2\kappa^2)\right)$  \\
    QPP~\cite{QPP}  & Purified Query & $\Tilde{\cO}\left(  1/(\varepsilon^2\kappa)\right)$ \\
    QSF (this work) & Identical Copies& $\Tilde{\cO}\left( \gamma^2/(\varepsilon^2\kappa)\right)$ \\
\bottomrule
\end{tabular}
\caption{Comparison of algorithms on estimating von Neumann entropy within additive error $\varepsilon$. Here $\Tilde{\cO}$ notation neglects $\log$ factors and $\kappa$ refers to the minimal nonzero eigenvalue of the quantum state, where we substitute $r$ with $\kappa = \mathcal{O}(1/r)$ to facilitate a more consistent comparison with analyses originally based on the matrix rank $r$. }
\label{tab: comparison entropy}
\end{table}

%%%%%%%%%%%%%%%%%%%%%%%%%%%%%%%%%%%%%%%%%%%%%%%%%%%%%%%%%%%%%%%%%%%%%%%%%%%%%%%%%%%%%%%%%%%%%
%%%%%%%%%%%%%%%%%%%%%%%%%%%%%%%%%%%%%%%%%%%%%%%%%%%%%%%%%%%%%%%%%%%%%%%%%%%%%%%%%%%%%%%%%%%%%
\section{Estimate Fidelity}\label{apx:fidelity}
Quantum state fidelity is a measure of the closeness between two quantum states. For two quantum states, $\rho, \sigma\in\mathbb{C}^{d\times d}$, the fidelity is given by $F(\rho, \sigma) = \left(\tr ( \sqrt{\sqrt{\rho} \sigma \sqrt{\rho}} ) \right)$. We can further take the fact that $\sqrt{\rho} \sigma \sqrt{\rho}$ shares same eigenvalues with $\rho\sigma$, to simplify it into $\tr ( \sqrt{\rho\sigma})$~\cite{baldwin2023efficiently}. Here, we directly extend the QSF framework to estimate fidelity between two quantum states using a polynomial expansion. To achieve that, we prepare $n$ copies of $\rho$ and $\sigma$, arranged alternately in the sequence $(\rho \ox \sigma )^{\ox n}$. Such a composite operation utilizes the cyclic-permutation-trick, satisfying
\begin{equation}\label{eq: fid}
    \tr[(\rho \ox \sigma )^{\ox n} \cdot P_{2k} ] = \tr[(\rho \sigma)^k], \quad k \leq n, 
\end{equation}
where $P_{2k}$ denotes the cyclic permutation of $2k$ quantum systems. Hence, any normalized polynomial approximation to $f(\rho \sigma)$ up to $n$-th order can be estimated via such an extension to Algorithm~\ref{alg:qsf}. 

% \begin{tcolorbox}
    \begin{corollary}\label{co:fidelity}
        For quantum states $\rho,\sigma\in\mathbb{C}^{d\times d}$ that the minimal nonzero eigenvalue of $\rho \sigma$ is $\kappa \in \mathbb{R}_+$, the QSF framework can approximate their fidelity $F(\rho, \sigma)$ with precision $\varepsilon$, required sample complexity $\Tilde{\cO}\left(\frac{\gamma^2}{\varepsilon^2\kappa}\right)$. 
    \end{corollary}
% \end{tcolorbox}
\begin{proof}
  Without loss of generality, suppose $\kappa, \varepsilon \in (0, \frac{1}{2}]$. According to Lemma 17 in~\cite{gilyen2022improved}, there is a polynomial $f_N(x)$ with an even/odd degree $N = \cO(\frac{1}{\kappa} \ln\frac{1}{\varepsilon})$ such that $|f_N(x) - \frac{1}{2} \sqrt{x}|\leq \varepsilon$. Denoting the rank of $\rho\sigma$ as $r$, there exists a polynomial function $f_N(\rho,\sigma)$ with respect to $\rho\sigma$, which is $\varepsilon$-close to the target function $F(\rho, \sigma)=\tr(\sqrt{\rho\sigma})$, i.e. $|f_N(\rho,\sigma) -\tr(\sqrt{\rho\sigma})| \leq \varepsilon/2$ with a sufficiently large $N = \cO(\frac{1}{\kappa} \ln\frac{r}{\varepsilon})$. Due to $\kappa = \mathcal{O}(\frac{1}{r})$, one can take $N = \cO(\frac{1}{\kappa} \ln\frac{1}{\varepsilon \kappa})$. Similar to the proof of Corollary~\ref{co:entropy}, Algorithm~\ref{alg:qsf} can estimate the normalized polynomial function $f_N(\rho,\sigma)/\gamma$ within precesion $\frac{\varepsilon}{2\gamma}$, i.e. $|\hat{f}-f_{N}(\rho,\sigma)/\gamma|\leq\frac{\varepsilon}{2\gamma}$, at a cost of $\cO(\frac{N\gamma^2}{\varepsilon^2})$ copies of $\rho$ and $\sigma$. Hence, we have $|\gamma\hat{f}-F(\rho,\sigma)|\leq|\gamma\hat{f}-f_N(\rho,\sigma)| + |f_N(\rho,\sigma)-F(\rho,\sigma)| \leq \varepsilon$. Since The total sample complexity is $\Tilde{\cO}(\frac{\gamma^2}{\varepsilon^2\kappa})$, where logarithmic factors have been omitted, which completes this proof.

%   Since $\kappa = \mathcal{O}(1/r)$, we obtain $N = \mathcal{O}\left(\frac{1}{\kappa} \ln \frac{1}{\varepsilon \kappa} \right)$. Following the same procedure as in Corollary~\ref{co:entropy}, Algorithm~\ref{alg:qsf} can estimate the normalized polynomial function $f_N(\rho, \sigma)/\gamma$ within precision $\varepsilon/(4\gamma)$, i.e.,
%     \[
%     \left|\hat{f} - \frac{f_N(\rho, \sigma)}{\gamma} \right| \leq \frac{\varepsilon}{4\gamma},
%     \]
%   at a cost of $\mathcal{O}(N \gamma^2 / \varepsilon^2)$ copies of $\rho$ and $\sigma$.

% By triangle inequality, it follows that
% \[
% \left| \gamma \hat{f} - \tr(\sqrt{\rho \sigma}) \right| 
% \leq \left| \gamma \hat{f} - f_N(\rho, \sigma) \right| + \left| f_N(\rho, \sigma) - \tr(\sqrt{\rho \sigma}) \right| \leq \varepsilon/2.
% \]
% Since both $\gamma \hat{f}$ and $\sqrt{F(\rho, \sigma)}$ lie in $[0, 1]$, we have
% \[
% \left| \gamma^2 \hat{f}^2 - F(\rho, \sigma) \right| \leq \left| \gamma \hat{f} - \sqrt{F(\rho, \sigma)} \right| \cdot \left| \gamma \hat{f} + \sqrt{F(\rho, \sigma)} \right| \leq \varepsilon.
% \]
% Hence, the total sample complexity is $\Tilde{\mathcal{O}}\left( \frac{\gamma^2}{\varepsilon^2 \kappa} \right)$, where logarithmic factors are omitted. This completes the proof.

\end{proof}

\begin{figure*}[htbp]
    \centering
    \includegraphics[width=0.9\textwidth]{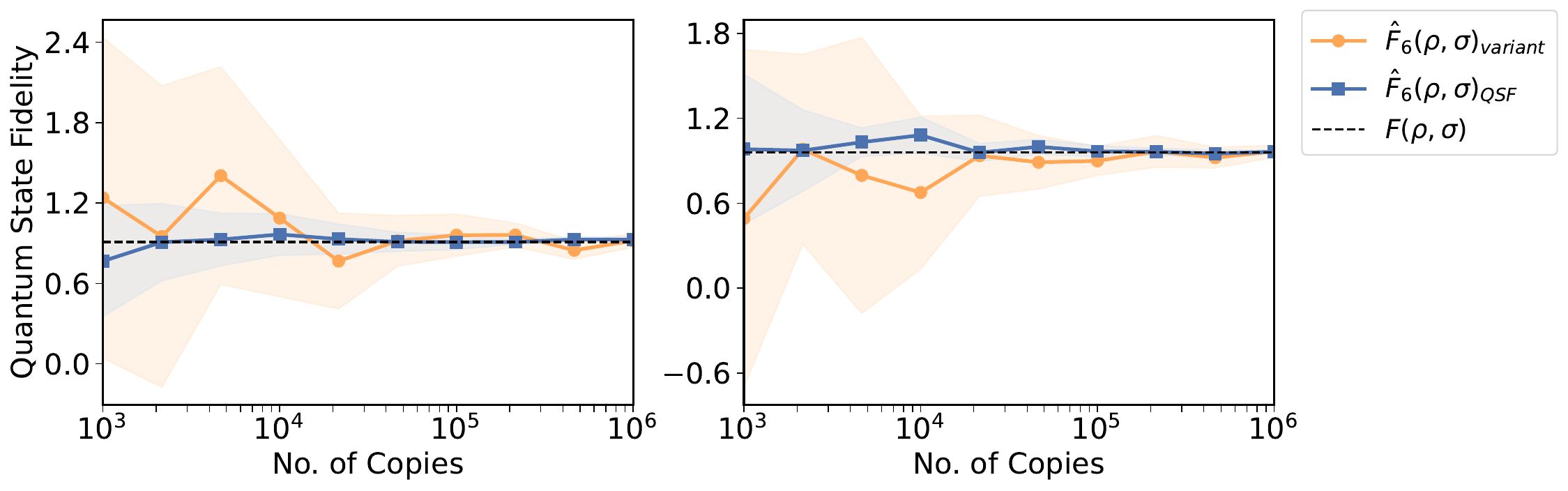}
    \caption{Numerical analysis of quantum state fidelity estimation. (a) Two random qubit states. (b) A random qubit state and the completely mixed state $I/2$. The dashed black line represents the actual fidelity value, denoted as $F(\rho, \sigma)$. The orange curves, labeled as $\hat{F}_6(\rho, \sigma)_{\text{variant}}$, represent the degree-6 average estimation obtained using the variant QSF method, while the blue curves, labeled as $\hat{F}_6(\rho, \sigma)_{\text{QSF}}$, represent the estimation using the standard QSF method. Each estimation is based on 10 repeats for trials, with the shadowed area representing the standard deviation.
    The specific quantum circuits are found in the previous chapter.}
    \label{fig:fidelity_numerical}
\end{figure*}

We also provide specific circuits for estimating quantum state fidelity through a degree-$6$ polynomial function $F_6(\rho, \sigma)$ in Figure.~\ref{fig:fidelity_circuit}. Numerical experiments for Algorithm~\ref{alg:qsf} and its variant to estimate the quantum state fidelity of two random quantum states and a random quantum state comparing with the maximally mixed state is shown in Fig~\ref{fig:fidelity_numerical}. The standard QSF also achieves better convergence than the variant as the estimation of von Neumann entropy. It is seen that the estimation converges to the fidelity at $10^5$ and $10^6$ copies for the standard and variant QSF, respectively. 

\begin{figure*}[htbp]
    \centering
    \resizebox{\textwidth}{!}{%
        \begin{quantikz}
            \lstick{$A^\prime\ \ \ \ \ \ \ \ \ \ \ \ket{0}$} & &  & & & & 
            & \gate{R_y(\frac{\pi}{2})} & \ctrl{1} & \gate{R_y(-\frac{\pi}{2})} & \ctrl{1} & \gate{R_y(\frac{\pi}{2})} & \ctrl{1} & \gate{R_y(-\frac{\pi}{2})} & \ctrl{1} & \gate{R_y(\frac{\pi}{2})} & \ctrl{1} & \gate{R_y(-\frac{\pi}{2})} & \ctrl{1} & \meter{X}\\
            \lstick[3]{$A\ \ \ \ \ \ \ket{0}^{\otimes 3}$} & & \gate{R_y(\beta_1)} \gategroup[3,steps=4,style={dashed,rounded corners,fill=blue!20, inner xsep=0pt},background,label style={label position=north east,anchor=north west}]{$\times 4$} & \ctrl{1} & & \targ{} &
            & \octrl{-1}  & \octrl{1}  & \octrl{-1}    & \octrl{1} & \octrl{-1}    & \octrl{1} & \octrl{-1}    & \octrl{1} & \ctrl{-1} & \ctrl{1}& \ctrl{-1} & \ctrl{1} & \qw\\
            &  & \gate{R_y(\beta_2)} & \targ{} & \ctrl{1} & &
            & \octrl{-1} & \octrl{1}   & \octrl{-1}    & \octrl{1} & \ctrl{-1}     & \ctrl{1}  & \ctrl{-1} & \ctrl{1}  & \octrl{-1} & \octrl{1}&\octrl{-1} & \octrl{1}& \qw\\
            &  & \gate{R_y(\beta_3)} & & \targ{} & \ctrl{-2} &
            & \octrl{-1} & \octrl{1}   & \ctrl{-1}     & \ctrl{1}  & \octrl{-1}    & \octrl{1} & \ctrl{-1} & \ctrl{1}    & \octrl{-1} & \octrl{1} & \ctrl{-1} & \ctrl{1}& \qw\\
            \lstick{$B\ (\rho \otimes \sigma)^{\otimes 6}$} & \qwbundle{} & & & & & 
            & & \gate{P_2} & & \gate{P_4} & & \gate{P_6} & & \gate{P_8} & & \gate{P_{10}} & & \gate{P_{12}} & \qw
        \end{quantikz}
    }
    \caption{Quantum circuit for calculating the state fidelity of two quantum states $\rho, \sigma\in\mathbb{C}^{d \times d}$, using the polynomial function $F_6(\rho, \sigma) = \sum_{j=0}^6 \alpha_j \tr\left(\left(\rho\sigma\right)^j\right)$, where $\alpha_j \in \mathbb{R}$. Notably, instead of being encoded into the PQC, the $0$-th order term is augmented directly. The purple section represents a predetermined PQC $V(\bm\beta)$ designed to generate a 3 qubit state and $P_j$, $j=2,4,6,8,10,12$, denotes the permutation operators.}
    \label{fig:fidelity_circuit}
\end{figure*}

Classical algorithms combined with measurements discussed the distance measure of two unknown mixed states~\cite{o2015quantum,buadescu2019quantum}, confronted with an exponential increase of resources with the scale of the quantum system, even if quantum states are restricted to be low-rank. Recent advancements in quantum algorithms~\cite{gilyen2022improved, wang2024new} have introduced QSVT-based protocols, and several variational quantum algorithms~\cite{cerezo2020variational,chen2021variational, tan2021variational} were implemented to estimate fidelity between two quantum states. 
\begin{table}[H]
\centering
\begin{tabular}{l|ccc}
\toprule
    \textbf{Algorithms}  & \textbf{Prerequisites} & \textbf{Sample Complexity} \\
\midrule
      State Certification~\cite{buadescu2019quantum}  & Identical Copies& $\Tilde{\cO}\left( d/\varepsilon\right)$  \\
     QSVT~\cite{wang2024new}  & Purified Query& $\Tilde{\cO}\left( 1/(\varepsilon^{7.5}\kappa^{6.5})\right)$  \\
     QSVT~\cite{gilyen2022improved}  & Purified Query& $\Tilde{\cO}\left( 1/(\varepsilon^5\kappa^{2.5})\right)$  \\

    QSF (this work) & Identical Copies& $\Tilde{\cO}\left( \gamma^2/(\varepsilon^2\kappa)\right)$ \\
\bottomrule
\end{tabular}
\caption{Comparison of algorithms on estimating fidelity of two $d$-dimensional quantum states within additive error $\varepsilon$. Here $\Tilde{\cO}$ notation neglects $\log$ factors and $\kappa$ refers to the minimal nonzero eigenvalue of the quantum state, where we substitute $r$ with $\kappa = \mathcal{O}(1/r)$ to facilitate a more consistent comparison with analyses originally based on the matrix rank $r$.}
\label{tab:comparison fid}
\end{table}

Table~\ref{tab:comparison fid} presents a comparison to those algorithms designed for estimating quantum fidelity, implying a significant reduction in the sample complexity required for estimating quantum fidelity with the QSF method. Quantum algorithms based on QSVT~\cite{gilyen2022improved, wang2024new} improved upon classical methods and circumvented the exponential resource overhead, but still rely on purified query access, which can be challenging to implement in practical settings and have copy complexities that scale unfavorably with $\varepsilon$ and $\kappa$. In contrast, QSF maintains the sample complexity of $\Tilde{\cO}\left( \frac{\gamma^2}{\varepsilon^2\kappa}\right)$ with a controllable factor $\gamma$ as estimating state fidelity. This makes it not only more accessible for current quantum technologies but also competitive in efficiency for measuring metrics with multivariate functions. The balance between practical implementation and resource scaling suggests that QSF is a strong candidate for applications involving the nonlinear function of quantum states, particularly in near-term quantum computing scenarios.

\end{document}